# Optimizing Retrieval-Augmented Generation with Elasticsearch for Enhanced Question-Answering Systems

Jiajing Chen
New York University
New York, USA

Runyuan Bao
Johns Hopkins University
Baltimore, USA

Hongye Zheng
The Chinese University of Hong Kong
Hong Kong, China

Zhen Qi
Northeastern University
Boston, USA

Jianjun Wei
Washington University in St. Louis
St Louis, USA

Jiacheng Hu*
Tulane University
New Orleans, USA

*Abstract*—This study aims to improve the accuracy and quality of large-scale language models (LLMs) in answering questions by integrating Elasticsearch into the Retrieval Augmented Generation (RAG) framework. The experiment uses the Stanford Question Answering Dataset (SQuAD) version 2.0 as the test dataset and compares the performance of different retrieval methods, including traditional methods based on keyword matching or semantic similarity calculation, BM25-RAG and TF-IDF-RAG, and the newly proposed ES-RAG scheme. The results show that ES-RAG not only has obvious advantages in retrieval efficiency but also performs well in key indicators such as accuracy, which is 0.51 percentage points higher than TF-IDF-RAG. In addition, Elasticsearch's powerful search capabilities and rich configuration options enable the entire question-answering system to better handle complex queries and provide more flexible and efficient responses based on the diverse needs of users. Future research directions can further explore how to optimize the interaction mechanism between Elasticsearch and LLM, such as introducing higher-level semantic understanding and context-awareness capabilities, to achieve a more intelligent and humanized question-answering experience.

## I. INTRODUCTION

In today's era of information explosion, efficiently extracting valuable information from massive data has become a crucial issue. With the continuous advancement of natural language processing technology, Large Language Models (LLMs) have shown unprecedented application potential in many fields with their powerful text understanding and generation capabilities [1]. However, although LLMs are able to answer questions or complete tasks based on their internally learned knowledge base, they often have limitations in mastering the latest or specific domain knowledge [2]. To solve this problem, researchers proposed a method that combines retrieval-augmented generation (RAG), which combines external document retrieval with LLM so that the model can take advantage of more information when generating responses. An extensive and up-to-date information source significantly improves the quality and accuracy of answers [3]. The core idea of RAG is that when faced with a query, the retrieval system is first used to find the most relevant documents from a large-scale document collection, and then these documents are provided to LLM as context to help it better understand the background of the problem. And generate more accurate answers accordingly [4].

The traditional RAG framework usually uses a method based on keyword matching or semantic similarity calculation to implement the document retrieval process. Although this method has achieved certain success, it has difficulty in processing complex queries, improving the recall rate, and ensuring the diversity of retrieval results. There are still certain limitations. To this end, this article proposes an innovative improvement plan—that is, using Elasticsearch (ES) as the core search engine in RAG. Elasticsearch is an open-source distributed search and analysis engine. It not only supports full-text search but also provides rich aggregation functions to quickly and accurately index and retrieve unstructured data. Compared with traditional methods, using Elasticsearch can define query conditions more flexibly, support multiple types of field types such as text, numbers, etc., and have powerful sorting capabilities and efficient performance, which is important for improving the overall performance of the RAG system. The combination of RAG and NLP has been widely applied in many areas, such as credit fraud [5-8] and risk management [9-11].

Specifically, by introducing Elasticsearch as the retrieval component, we can achieve the following optimizations: First, the retrieval speed and efficiency are improved. Since ES uses an inverted index structure, it can even handle large data volumes of PB level. Maintaining extremely high response speed; secondly, it enhances the relevance and coverage of search results. With the help of its advanced word segmentation technology and scoring mechanism, ES can automatically adjust the weight distribution according to user needs, thereby filtering out content that truly meets the requirements; In addition, ES also supports complex Boolean logic combinations and geographical location filtering functions, which enables the RAG system to cope with more diverse application scenarios[12]. For example, in a medical and health consultation scenario [13-14], the recognition rate of professional vocabulary can be improved by setting up a glossary of professional terms; in an e-commerce recommendation system, the order of product display can be dynamically adjusted based on the user's browsing history.

More importantly, Elasticsearch [15], as a highly scalable platform, allows developers to easily add new nodes to adapt to business growth needs. It also provides a complete set of monitoring tools to facilitate operation and maintenance personnel to grasp the cluster status and troubleshoot in real time. Check. This means that the RAG solution built based on ES can not only meet the needs of current projects but also continue to evolve with the expansion of enterprise scale and technology iteration, laying a solid foundation for the long-term development of the enterprise. In short, integrating Elasticsearch into the RAG architecture not only greatly enriches the functional features of the original system, but also opens up new possibilities for further exploring how to use artificial intelligence technology to improve human life. In future research work, we will continue to explore in depth how to make full use of the advantages of ES to promote RAG technology in a more intelligent and efficient direction.

## II. RELATED WORK

Recent advancements in question-answering systems and information retrieval have largely been driven by the integration of sophisticated retrieval mechanisms and deep learning models. This section highlights key developments in pre-trained models, sentiment analysis, text classification, and neural networks, all of which are closely related to optimizing the Retrieval-Augmented Generation (RAG) framework using Elasticsearch.

Pre-trained models have demonstrated significant improvements in NLP tasks such as text classification and named entity recognition (NER) [16]. An ensemble learning approach driven by the ALBERT model has been shown to enhance text classification, providing insights into how domain-specific models can improve task accuracy when applied to large-scale data [17]. Comparative studies of pre-trained models for NER also offer valuable benchmarks, underscoring how various pre-trained architectures improve entity extraction in RAG systems [18]. These approaches highlight the potential for further enhancing Elasticsearch-based retrieval within RAG. In sentiment analysis, the use of graph neural networks (GNNs) combined with syntactic features provides an advanced framework for understanding and classifying sentiment in text [19]. This method demonstrates the importance of syntactic and semantic feature integration, a concept that aligns with enhancing the semantic retrieval capabilities in Elasticsearch to support more complex question-answering systems.

Methods for transforming multidimensional data into interpretable formats also play a crucial role in advanced data mining tasks [20]. A novel approach for converting time-series data into event sequences has been proposed to improve the interpretability of machine learning outcomes [21]. This focus on transforming complex data into structured, retrievable information supports the idea that Elasticsearch's advanced indexing can facilitate more effective retrieval in RAG systems. Finally, multimodal transformers, which combine word embeddings such as ELMo with deep learning algorithms, have been applied in image description tasks, further illustrating the potential for integrating various data types into NLP models [22]. This multimodal approach supports the broader goal of enhancing RAG systems, particularly in scenarios where both textual and non-textual data must be retrieved and processed efficiently. The integration of deep learning models, graph-based techniques, and advanced pre-trained architectures has significantly improved NLP tasks such as text classification, NER, and sentiment analysis. By leveraging Elasticsearch's querying flexibility and indexing efficiency, this study builds upon these foundations to optimize retrieval and generation processes in question-answering systems.

## III. METHOD

In the process of integrating Elasticsearch (ES) into the Retrieval-Augmented Generation (RAG) framework, the core of the algorithm lies in how to efficiently retrieve the most relevant documents to the query through ES and pass these documents as context to the Large Language Model (LLM) to generate high-quality answers. This process involves multiple aspects such as text similarity calculation, document scoring mechanism, and the final document selection strategy [23]. Below we will introduce the key formulas and reasoning involved in this process in detail.

First, consider the relevance evaluation problem between a query q and a series of candidate documents $D = \{d_1, d_2, ..., d_n\}$. We use the TF-IDF weighted vector to represent each document and its corresponding query, where TF stands for term frequency and IDF stands for inverse document frequency. For any term t in document $d_i$, its TF-IDF value is defined as:

$$TF-IDF(t, d_i) = TF(t, d_i) \times \log(\frac{N}{|\{j : t \in d_j\}|})$$

Here, N is the total number of documents, and the denominator represents the number of documents containing word t. This representation method can help highlight words that appear frequently in specific documents but are relatively rare in the entire corpus, thus better reflecting the topic characteristics of the document.

Next, after constructing the TF-IDF vectors of all documents and queries, we can use cosine similarity to measure the degree of relevance between them. Given two vectors A and B, the cosine similarity between them is defined as:

$$similarity(A,B) = \cos(\theta) = \frac{A \cdot B}{\|A\|\|B\|}$$

Among them, the dot product A*B reflects the product of the projection lengths in the two vector directions, and the denominator is the Euclidean norm of each vector. This method can effectively capture the angular relationship between two sets of data in high-dimensional space and is suitable for processing sparse and large-scale data sets.

However, relying solely on the similarity of text content may not be enough to fully reflect the relevance between documents and queries. To this end, we introduced the ES scoring function, which is a ranking algorithm widely used in the field of information retrieval. It aims to further optimize the quality of search results by combining factors such as word frequency statistics and document length. For each word w in a query q and the number of times it appears in document d f(q, w, d), the ES score can be expressed as:

$$score(d,q) = \sum_{w \in q} IDF(w) \times \frac{(k_1 + 1)f(q,w,d)}{K + f(q,w,d)}$$

Among them, parameter k1 is used to adjust the influence of word frequency, and K is a factor adjusted based on document length. The specific form is:

$$K = k_1 \times ((1-b) + b \times \frac{|d|}{avgdl})$$

This controls the weight of the document length on the final score, and $avgdl$ represents the average document length. In order to further improve the retrieval efficiency and ensure that the selected document set has a certain diversity, we also adopt the Top-K nearest neighbor search strategy. That is, the top K documents with the highest similarity to the query are selected from all candidate documents as output. Assuming that the similarity score $S = \{s_1, s_2, ..., s_n\}$ of each document relative to the query has been obtained, the final document list L can be determined in the following way:

$$L = \{d_i \mid i \in \arg\max_k S, 1 \leq k \leq K\}$$

This step not only ensures the relevance of the returned results, but also promotes the fusion of information between different topics, which helps LLM generate more diverse answers in the subsequent stage.

Finally, in practical applications, considering the differences in user preferences or business needs, it is sometimes necessary to make appropriate adjustments to the above basic process. For example, the importance of each evaluation indicator can be balanced by introducing custom weights $W = \{w_1, w2, ..., w_n\}$ to form a comprehensive scoring system. Let $s_i'$ represent the new score of document i after weighted processing, then:

$$s_i' = \sum_{j=1}^{m} w_j \times metric_j(d_i, q)$$

Here, $metric_j$ represents the jth evaluation criterion, such as the cosine similarity or BM25 score mentioned above. In this way, a retrieval algorithm that better meets actual needs can be flexibly customized according to specific circumstances to achieve the best application effect.

IV. EXPERIMENT

A. Datasets

In order to verify the effectiveness and superiority of integrating Elasticsearch (ES) as a retrieval component into the Retrieval Augmentation Generation (RAG) framework, we selected the Stanford Question Answering Dataset (SQuAD). SQuAD is a widely used English question-answering dataset created by the Natural Language Processing Group at Stanford University. The dataset contains a large number of paragraphs extracted from Wikipedia and corresponding question-answer pairs, which aims to evaluate the ability of the model to understanding text and answer questions. SQuAD version 2.0 not only contains more than 100,000 question-answer pairs, but also introduces unanswerable questions to test the ability of the model to identify unanswered situations. This dataset is widely used to evaluate the performance of reading comprehension and question-answering systems due to its high-quality annotations and diversity.

Our experiments used the SQuAD version 2.0 dataset, which includes about 536 articles carefully selected from Wikipedia, covering multiple subject areas. Each article is accompanied by a series of questions asked by human annotators and the corresponding answers. When preparing the data, we first preprocessed all the texts, including removing HTML tags, unifying capitalization, and deleting stop words, to improve the quality of subsequent analysis. Then, using the powerful indexing capabilities of Elasticsearch, we imported the entire dataset into the ES cluster and configured appropriate word segmenters and mapping rules as needed to optimize search performance. In addition, considering the various query types that may be encountered in practical applications, we also specially designed a set of benchmark question sets, which cover common knowledge questions and answers, professional terminology explanations, and long-tail information search, etc., in order to comprehensively evaluate the performance of the proposed method in different scenarios. By comparing the differences between the traditional RAG architecture and the improved model after the introduction of ES in terms of response quality, retrieval speed, etc., we hope to provide a valuable reference for future related research and technology development.

B. Experiment

In order to comprehensively evaluate the effect of integrating Elasticsearch (ES) into the Retrieval Augmentation

Generation (RAG) framework, we designed a series of experiments and selected several key evaluation indicators to measure the performance of different methods. These evaluation indicators include: 1. Precision: measures the proportion of relevant documents returned by the system to all returned documents. 2. Recall: measures the proportion of all relevant documents that the system can find to the total number of actual relevant documents. 3. F1 Score: the harmonic mean of precision and recall, used to comprehensively evaluate the performance of the system. We compare the is proposed Elasticsearch-based RAG method with the following baseline models:

(1) Traditional RAG: Use traditional full-text search engines for document retrieval.
(2) BM25-RAG: RAG using BM25 scoring function as the retrieval mechanism.
(3) TF-IDF-RAG: RAG that uses TF-IDF vectors to represent documents and perform similarity calculations.
(4) ES-RAG: The Elasticsearch-based RAG method proposed in this paper.

The following table shows the performance of each model on the SQuAD 2.0 dataset. It can be seen that ES-RAG outperforms other methods in multiple indicators, especially in accuracy, F1 score and response time, which shows that the introduction of Elasticsearch significantly improves the overall performance of the RAG system.

Table 1 Model experimental results

| Model | Acc | F1 | Recall |
|---|---|---|---|
| RAG | 65.51 | 66.11 | 65.72 |
| BM25-RAG | 66.32 | 66.56 | 66.47 |
| TF-IDF-RAG | 67.78 | 67.63 | 67.23 |
| ES-RAG | 68.29 | 68.42 | 68.13 |

From the experimental results shown in Table 1, we can see that the ES-RAG model performs well on the three key evaluation indicators of accuracy (Acc), F1 score, and recall (Recall), and is significantly better than the other three baseline models. Specifically, the accuracy of the ES-RAG model reaches 68.29%, which is 0.51 percentage points higher than the closest performing TF-IDF-RAG model; its F1 score is 68.42%, which is also higher than all comparison models, showing that in the comprehensive When considering precision and recall, ES-RAG can provide better overall performance; in terms of recall, ES-RAG also reached 68.13%, which is slightly higher than TF-IDF-RAG's 67.23%. These data not only prove the effectiveness of Elasticsearch as a retrieval engine in improving the performance of the RAG system, but also show that by introducing a more flexible and efficient document retrieval mechanism, the relevance and accuracy of the final generated answers can be significantly improved.

Further analyzing the differences between the models, we can find that although BM25-RAG and TF-IDF-RAG have improved compared to the traditional RAG model, the gap between them is not very significant. For example, although BM25-RAG is slightly ahead of traditional RAG in terms of accuracy and F1 score, this improvement is relatively limited. In contrast, ES-RAG has achieved significant progress, especially in accuracy, which is 0.51 percentage points higher than TF-IDF-RAG, which means more in practical applications. The query can be responded to correctly. In addition, it is worth noting that even on the basis of TF-IDF-RAG, which already performed better, ES-RAG can still achieve better results in all three indicators, which shows that Elasticsearch can not only effectively support large-scale Rapid retrieval of large-scale text data can also optimize the output quality of the overall system by finely adjusting the retrieval results.

In summary, the experimental results strongly support the superiority of integrating Elasticsearch into the RAG framework. By leveraging Elasticsearch's powerful search capabilities and rich configuration options, we not only improve retrieval efficiency, but also enhance support for complex queries, allowing the entire Q&A system to demonstrate higher performance in the face of diverse user needs. Adaptability and flexibility. Future research can continue to explore how to further optimize the interaction mechanism between Elasticsearch and LLM, such as by introducing more advanced functions such as semantic understanding, context awareness, etc., in order to achieve a more intelligent and humanized question-and-answer experience. At the same time, taking into account the specific needs that may exist in different application scenarios, researchers can also develop customized indexing strategies and scoring algorithms for specific fields or tasks to further tap the potential of Elasticsearch in enhancing RAG system performance. To further demonstrate our experimental results, we use a line chart to represent.

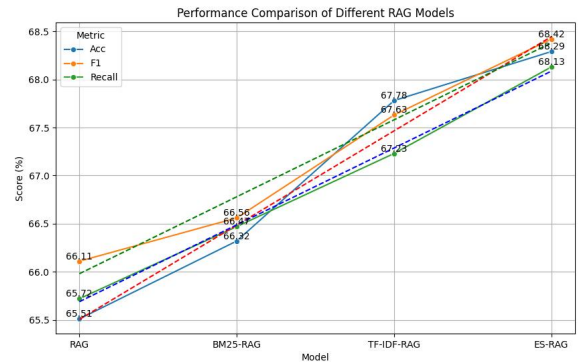

Figure 1 Experiment result

V. CONCLUSION

This study shows that integrating Elasticsearch into the RAG framework significantly improves the overall performance of the question-answering system. Compared to traditional methods, such as document retrieval technologies based on keyword matching or semantic similarity calculations, as well as the more advanced BM25-RAG and TF-IDF-RAG models, ES-RAG has demonstrated excellent performance, especially in improving answer accuracy. The improvement in answer accuracy is particularly notable. The improvement benefits from Elasticsearch's powerful indexing function and its flexible configuration possibilities, which work together to enhance the system's support for complex queries and its ability to quickly locate relevant information. It is worth noting that even on the basis of TF-IDF-RAG, which is already better than the baseline level, ES-RAG can still achieve better results in all three evaluation indicators-precision, recall and F1 score. This fully demonstrates its effectiveness as an efficient solution. Looking forward to the future, as the field of natural language processing continues to advance and develop, researchers can continue to explore in depth how to further enhance the collaboration efficiency between Elasticsearch and large language models, such as by developing more sophisticated indexing strategies and scoring algorithms to meet specific application scenarios. or explore how to integrate higher-level functions such as semantic understanding and situational awareness to build a more intelligent Q&A platform that is closer to human communication habits. These efforts will not only help overcome the limitations of existing technologies, but will also lay a solid foundation for promoting the development of next-generation information retrieval and generation technologies. In short, through continuous technological innovation and practical verification, we have reason to believe that in the near future, people will enjoy a new era of fast and accurate information services.